\documentclass{aa} 

\usepackage{amssymb, amsmath}
\usepackage{amsmath}
\usepackage{graphicx} 
\usepackage{txfonts} 
\usepackage{natbib, hyperref} 
\usepackage{subcaption}         % necessary for continued figures 
\usepackage{lscape}             % to rotate a single page table, example in appendix.
\usepackage{placeins}           % useful with \FloatBarrier, to keep 
\begin{document} 

\title{Gas Accretion versus BH Merger driven Growth Modes of Supermassive Black Holes 
and Implications for the Little Red Dots} 

\titlerunning{Accretion versus Merger Growth Modes of SMBHs} 

% Central Supermassive Black Holes and Implications for the Little Red Dots 
% Assembly ... and Host Galaxies ... along the $M_{\rm BH} - M_{\star}$ Relation

\author{Paramita Barai \inst{1} \inst{2} \inst{3} \thanks{paramita.barai@inaf.it}}

\institute{
$^1$ Istituto Nazionale di Astrofisica (INAF) - Osservatorio Astronomico di Trieste (OATs), 
Via G.B. Tiepolo 11, I-34143 Trieste, Italy \\ 
$^2$Centro de Ci\^{e}ncias Naturais e Humanas - Universidade Federal do ABC, 
Av. dos Estados 5001, Santo Andr\'{e} - SP, 09210-580, Brazil \\ 
$^3$N\'{u}cleo de Astrof\'{i}sica - Universidade Cidade de S\~{a}o Paulo, 
Rua Galv\~{a}o Bueno 868, S\~{a}o Paulo - SP, 01506-000, Brazil 
} 

% \date{Received September 22, 2025}

\abstract
{
We investigate the growth of central supermassive black holes in galaxies, 
aiming to distinguish between gas accretion versus BH merger-driven growth modes. 
By performing and analysing cosmological hydrodynamical simulations of $(50 ~ {\rm Mpc})^3$ 
comoving boxes, we also study how the BH feedback parameters 
affect the coevolution between SMBHs and their host galaxies. 
Starting as $10^5 M_{\odot}$ seeds, 
we find that the BHs grow initially via BH mergers to $\sim 10^7 M_{\odot}$. 
Gas accretion onto the BHs is initially low, 
then increases with time, and reaches the Eddington rate after $7-9$ Gyrs. 
The BHs then undergo very fast growth via efficient gas accretion 
over a period of $600 - 700$ Myr, when the BH mass increases $10^2 - 10^3$ times, 
causing their predominant growth from $10^7 M_{\odot}$ to $(10^9 - 10^{10}) M_{\odot}$. 
Taking into account the cosmological gas inflows and outflows, 
SMBHs do not grow to more than $10^{10} M_{\odot}$ by $z=0$, 
because of gas depletion from galaxy centers driven by AGN feedback. 
In terms of SMBH - host galaxy coevolution along the $M_{\rm BH} - M_{\star}$ relation, 
we find that they initially lie below and thereby move upward toward the relation. 
We make some physical implications of the growth of high-$z$ Little Red Dots recently observed by JWST: 
the normal-mass SMBHs had predominantly undergone BH merger driven evolution, 
whereas the overmassive BHs underwent periods of 
Eddington-limited or super-Eddington bursts of gas accretion. 
} 

\keywords{cosmological hydrodynamical simulations -- supermassive black holes -- black hole galaxy coevolution} 

\maketitle

\nolinenumbers

%  The star formation rate density (SFRD) evolution of the DGs 
% (stellar mass $10^{5} - 10^{8} M_{\odot}$) has a peak plateau between $z = 4 - 6$. 
% The SFRD is reduced by factors up to $3$, when the BHs have grown to a few times $10^5 M_{\odot}$. 

\section{Introduction} 

Supermassive Black Holes (SMBHs) exist at the centers of active galactic nuclei (AGN), which liberate 
enormous amounts of feedback energy powered by the accretion of matter \citep[e.g.,][]{Rees84, Urry95}. 
AGN are widely observed via multi-wavelength observations, 
starting from the local Universe up to $13.2$ Gyr ago \citep[e.g.,][]{Fan2023, Natarajan2024}. 
SMBHs already grown to mass $\geq 10^9 M_{\odot}$ are observed in luminous quasars at $z \sim 6$, 
when the Universe was less than $1$ Gyr old \citep[e.g.,][]{Willott03, Yang2023}. 

How the SMBHs in AGN grew to billions of solar masses involves unresolved questions, 
despite progress in understanding the SMBH accretion 
and feedback on the environment \citep[e.g.,][]{Alexander2025}. 
Especially at early epochs, it is difficult to understand how they formed over such short time-scales, 
and there are open issues with various plausible scenarios \citep[e.g.,][]{Dijkstra08, Inayoshi12, Matsumoto15}. 
Some studies advocate growth from stellar-mass BHs by rapid enhanced super-Eddington accretion, 
formation of direct-collapse massive $10^5 M_{\odot}$ BH seeds soon after the Big Bang \citep[e.g.,][]{Volonteri2005}, 
via mergers of intermediate-mass black holes \citep[e.g.,][]{Barai2019}. 

Adding to the population of early SMBHs, recent JWST observations are 
revealing $10^{7} - 10^{8} M_{\odot}$ BHs at $z \sim 8-12$ 
\citep[e.g.,][]{Kokorev2023, Kocevski2025, Bhatt2024, Scholtz2025}. 
Some of these first SMBHs are overmassive in their host galaxy 
with respect to the $z=0$ SMBH-to-stellar mass correlation 
\citep[e.g.,][]{Goulding2023, Maiolino2024, Wu2025}. 
At the same time, a population of normal-mass central SMBHs 
(with the mean SMBH-to-stellar mass ratio of $\sim 0.1\%$, consistent with the local relation) 
have been detected in $z \sim 3-6$ AGN \citep[e.g.,][]{Li2025, Geris2025} observed by JWST. 

The formation channels of these "normal-mass" and overmassive SMBHs, 
in terms heavy seeds or super-Eddington accretion onto lighter stellar-mass seeds, 
have not been identified yet \citep[e.g.][]{Jeon2025}. 
The very existence of such normal SMBHs together with the overmassive population, 
imply that there might be diverse pathways for SMBH formation. 

Using SDSS/Subaru data, \citet{Li2021} found no significant evolution 
of the $M_{\rm BH} - M_{\star}$ relation of quasars between $z = 0.8-0.2$, 
and it remains consistent with the local relation. 
\citet{Graham2025} argue that the $M_{\rm BH} / M_{\star}$ correlation 
that a population of galaxies (including the JWST detected high-$z$ little red dots) follows, 
is dependent on galaxy morphology. 
Also using JWST observations, \citet{Juodzbalis2025} detected low-luminosity AGN 
at $z = 1.5-9$, accreting at likely sub-Eddington ratios, 
hosted in low mass ($M_{\star} \sim 10^{8} M_{\odot}$) galaxies; where the SMBHs are 
overmassive relative to the local $[M_{\rm BH} - M_{\star}]$ relation, 
while consistent with the local $[M_{\rm BH} - \sigma_{\star}]$ relation. 

Numerical simulations have advanced in parallel to understand these SMBH evolution. 
Concordance galaxy formation models based on cold dark-matter cosmology
widely invoke AGN feedback as a crucial ingredient to self-regulate galaxy and SMBH growth.
This has been studied in numerical hydrodynamical simulations 
\citep[e.g.,][]{DiMatteo05, Ostriker10, Barai11, Khandai2015, Dubois2016}, 
as well as semi-analytical models \citep[e.g.,][]{Kauffmann00, Somerville2008, Fontanot2020}. 

Related to the goals of our study, 
\citet{Habouzit2022} showed that the Illustris, TNG100, TNG300, Horizon-AGN, EAGLE, and SIMBA 
cosmological simulations do not agree on whether BHs at $z > 4$ are overmassive or undermassive 
% at fixed galaxy stellar mass 
with respect to the $M_{\rm BH} - M_{\star}$ scaling relation at $z = 0$. 
\citet{Haidar2022} investigated AGN populations in the 
Illustris, TNG, Horizon-AGN, EAGLE, and SIMBA simulations 
compared with current observational constraints in low-mass galaxies, 
finding that some simulations produce BHs that are too massive. 
Using $(18 ~ {\rm Mpc})^3$ simulations, \citet{Kho2025} found that different 
BH seeding models lead to different normalizations of the $M_{\rm BH} - \sigma$ relation. 
% About the BH growth mode, 
\citet{Kho2025} also found that the BH growth is merger-dominated 
in low-mass $(\leq 10^{9} M_{\odot})$ galaxies, 
and is accretion dominated in high-mass $(\geq 10^{9} M_{\odot})$ galaxies; 
which directly influences the $M_{\rm BH} - \sigma$ evolution. 

In our study, we investigate the growth of central SMBHs versus their host galaxies, 
by performing $(50 ~ {\rm Mpc})^3$ cosmological hydrodynamical simulations, 
aiming to distinguish between gas accretion versus BH merger-driven growth modes. 
We probe the SMBH-galaxy coevolution, 
in particular that of the black hole mass - stellar mass correlation. 
We also shed light on the growth modes of the overmassive versus normal-mass SMBHs 
in high-$z$ little red dots recently observed by JWST. 

% In particular, to distinguish between the growth modes of the overmassive and normal-mass SMBHs; 
% and identify if their feedback signatures are different and could be observable.
% The simulations will try to reproduce the overmassive and normal-mass SMBHs, 
% and determine the distinguished physical conditions (e.g. BH seed mass, 
% super-Eddington BH accretion, mergers, thermal/kinetic AGN feedback, environment) 
% required for the formation of each population. 
% The simulation analyses would quantify the feedback of the SMBHs and identify possible outflow signatures. 

% Cosmic Noon is the period of peak in the SFRD in the Universe $z \sim 2 - 3$ 
% when most galaxies have high star formation activities. 

% Galaxy Formation Efficiency (GFE), also called the Star formation efficiency, 
% is the ratio of stellar mass to halo mass of a galaxy. 
% Using the observational age distribution of stars in post-starburst galaxies (measured by HST photometry 
% assuming no dust extinction) of a galaxy cluster at $z \sim 2$ and the standard Press Schechter formalism, 
% \citet{Liu20} inferred a rapid evolution of the GFE from $10^{-4}$ to $0.01$ during $z \sim 20 - 13$. 

\section{Methodology} 
\label{sec-numerical} 

We perform $(50 ~ {\rm Mpc})^3$ cosmological hydrodynamical simulations, 
using a modified version of the code {\sc GADGET-3} \citep{Springel05}. 
The code uses the Tree-PM (particle mesh) and SPH (smoothed particle hydrodynamics) methods. 
The baryonic physical processes occurring in the multiphase interstellar medium (ISM), 
on scales unresolved in cosmological simulations, is modeled using spatially averaged 
properties describing the medium on scales that are resolved, 
as described in \S\ref{sec-num-cool-SF-SN} and \S\ref{sec-num-BH}. 
Our different simulation runs are outlined in \S\ref{sec-simulation}. 

\subsection{Cooling, Star-Formation, SN Feedback} 
\label{sec-num-cool-SF-SN} 

Radiative cooling and heating is implemented, including metal-line cooling \citep{Wiersma09a}. 
In this model, net cooling rates are computed element-by-element 
tracking $11$ atomic species: H, He, C, Ca, O, N, Ne, Mg, S, Si, Fe. 
A spatially-uniform time-dependent photoionizing background radiation is 
considered from the cosmic microwave background and the 
\citet{Haardt01} model for the ultraviolet/X-ray background. 
The gas is assumed to be dust free, optically thin, 
and in (photo-) ionization equilibrium. 
Contributions from the $11$ elements are interpolated as a function of density, 
temperature and redshift from tables that have been pre-computed using the 
public photoionization code CLOUDY \citep[last described by][]{Ferland98}. 

Star formation (SF) is adopted following the multiphase 
effective sub-resolution model of \citet{SH03}. 
Gas particles with density above a limiting threshold, 
$\rho_{\rm SF} = 0.13$ cm$^{-3}$ (in units of number density of hydrogen atoms), 
represent cold and hot phase regions of the ISM. 
Stellar evolution and chemical enrichment are computed for the 11 elements 
\citep{Tornatore07}. 
Each star particle is treated as a simple stellar population (SSP). 
Given a stellar initial mass function (IMF), the mass of the SSP is varied in time 
following the death of stars, and accounting for stellar mass losses. 
A fixed stellar initial mass function \citep{Chabrier03} is included, 
in the mass range $(0.1 - 100) M_{\odot}$. 
Stars within a mass interval $[8 - 40] M_{\odot}$ become SN first before 
turning into stellar-mass black holes at the end of their lives, 
while stars of mass $> 40 M_{\odot}$ are allowed to 
directly end in black holes without contributing to gas enrichment. 

Feedback from supernovae is incorporated in the kinetic form, 
assuming a mass ejection rate $\dot{M}_{\rm SN}$ 
proportional to the star formation rate ($\dot{M}_{\star}$): 
\begin{equation} 
\dot{M}_{\rm SN} = \eta \dot{M}_{\star} . 
\end{equation} 
The mass loading factor of SN wind is taken as $\eta = 2$ 
\citep[e.g.,][]{Tornatore07, Barai13, Melioli13}, 
following observations revealing that SN-driven outflow rates are comparable to 
or larger than SF rates of galaxies \citep[e.g.,][]{Martin99, Bouche12}. 
The SN wind kinetic power is a fixed fraction $\chi$ of SN internal energy rate: 
\begin{equation} 
\label{eq-SN-feedback} 
\frac{1}{2} \dot{M}_{\rm SN} v_{\rm SN}^2 = \chi \epsilon_{SN} \dot{M}_{\star} . 
\end{equation} 
Here $v_{\rm SN}$ is the SN wind velocity, 
$\epsilon_{SN}$ is the average energy released by SN for each $M_{\odot}$ 
of stars formed under the instantaneous recycling approximation. 
For our adopted \citet{Chabrier03} power-law IMF, 
$\epsilon_{SN} = 1.1 \times 10^{49}$ erg $M_{\odot}^{-1}$. 
Combining above expressions, $v_{\rm SN}$ can be re-written as: 
$v_{\rm SN} = \left( 2 \chi \epsilon_{SN} / \eta \right)^{1/2}$. 
Following a series of studies \citep[e.g.,][]{Tornatore07, Tescari11, Barai13}, 
and unlike \citep{SH03}, we choose $v_{\rm SN}$ as a free parameter. 
We adopt a constant-velocity outflow with SN wind velocity 
$v_{\rm SN} = 350$ km/s \citep[as was done in e.g.][]{Tornatore07, Barai15, Biffi16}. 
Bondi accretion boost factor, $\alpha_{\rm B} = 0.4$. 

%%%%%%%%%%%%%%%%%%%%%%%%%%%%%%%%%%%%%%%%%%%%%%%%%%%%%%%%%
%%%%%%%%%%%%%%%%%%%%%%%%%%%%%%%%%%%%%%%%%%%%%%%%%%%%%%%%%
% 
% TABLE 1 

\begin{table*} 
\caption{Simulation runs and parameter values changed in each case.} 
\label{Table-Sims} 
\begin{tabular}{@{}ccccccc} 

\hline 

Run  & BH      & Seed BH mass,                & BH feedback             & Type of     & BH kinetic feedback \\ 
name & present & $M_{\rm BHseed} [M_{\odot}]$ & efficiency $\epsilon_f$ & BH feedback & kick velocity $v_w$ (km/s) \\ 

\hline 

{\it SFonly} & -- & -- & -- & -- \\ % 33.87Mpc-N256-SN 

{\it SN} & No & -- & -- & -- \\ % 33.87Mpc-N256-SN 

{\it BHs5e0.1v5} & Yes & $10^{5}$ & $0.1$ & Kinetic & $5000$ \\ % (BHstd) 33.87Mpc-N256-SN-BH-6-TimeBetFoF1.4-alphaBondi0.4 

{\it BHs6e0.1v5} & Yes & $10^{6}$ & $0.1$ & Kinetic & $5000$ \\ % (BHabove) SEED ABOVE THE M_BH - M_STAR RELATION. 

{\it BHs4e0.1v5} & Yes & $10^{4}$ & $0.1$ & Kinetic & $5000$ \\ % (BHbelow)

{\it BHs5e1v5} & Yes & $10^{5}$ & $1$ & Kinetic & $5000$ \\ % (BHhiFeed) 33.87Mpc-N256-SN-BH-6-variant-3-epsF1.0 

{\it BHs5e0v5} & Yes & $10^{5}$ & $0$ & Kinetic & $5000$ \\ % (BHnoFeed)

{\it BHs5e0.1v10} & Yes & $10^{5}$ & $0.1$ & Kinetic & $10000$ \\ % (BHhiKick)

{\it BHs5e0.1v1} & Yes & $10^{5}$ & $0.1$ & Kinetic & $1000$ \\ % (BHloKick) 

{\it BHs5e0.1TRML} & Yes & $10^{5}$ & $0.1$ & Thermal & -- \\ % 33.87Mpc-N256-SN-BH-6-TimeBetFoF1.4-THERMALfeed

\hline 
\end{tabular} 
\end{table*} 

%%%%%%%%%%%%%%%%%%%%%%%%%%%%%%%%%%%%%%%%%%%%%%%%%%%%%%%%%
%%%%%%%%%%%%%%%%%%%%%%%%%%%%%%%%%%%%%%%%%%%%%%%%%%%%%%%%% 

\subsection{BH Accretion and Feedback} 
\label{sec-num-BH} 

We identify galaxies in our simulations by executing the FOF 
at time intervals of a multiplicative factor $1.4$ of the cosmological scale factor $a$, 
or, $a_{\rm next} / a_{\rm prev} = 1.4$. 
Massive galaxies with (i) a total halo mass higher than $10^{10} M_{\odot}$, 
(ii) a stellar mass higher than $5 \times 10^{7} M_{\odot}$, 
(iii) the gas mass is equal or larger than 10 percent of stellar mass, 
(iv) not containing a BH yet, are selected. 
A BH of initial mass $M_{\rm BH} = 10^{5} M_{\odot}$ is seeded at the center of each such massive halo. 

Gas is considered to accrete onto a BH according to the Bondi-Hoyle-Lyttleton 
accretion rate \citep[$\dot{M}_{\rm Bondi}$:][]{Bondi52}, 
\begin{equation} 
\label{eq-Mdot-Bondi} 
\dot{M}_{\rm Bondi} = \alpha_{\rm B} \frac{4 \pi G^2 M_{\rm BH}^2 \rho}{ \left(c_{s}^2 + v^2\right) ^ {3/2}} , 
\end{equation} 
where $G$ is the gravitational constant, $c_{s}$ is the sound speed, 
$\rho$ is the gas density, $v$ is the velocity of the BH relative to the gas, and 
$\alpha = 100$ is a numerical boost factor \citep[e.g.,][]{SDH05, Johansson09a, Dubois13}. 
Furthermore, accretion is limited to the Eddington mass accretion rate $(\dot{M}_{\rm Edd})$: 
$\dot{M}_{\rm BH} = {\rm min} \left( \dot{M}_{\rm Bondi}, \dot{M}_{\rm Edd} \right)$. 
The Eddington luminosity is used to express the Eddington mass accretion rate, 
\begin{equation} 
\label{eq-LEdd} 
L_{\rm Edd} = \frac{4 \pi G M_{\rm BH} m_p c} {\sigma_T} = \epsilon_r \dot{M}_{\rm Edd} c^2 , 
\end{equation} 
where $m_p$ is the mass of a proton, $c$ is the speed of light, 
and $\sigma_T$ is the Thomson scattering cross-section for an electron. 

Feedback energy is distributed to the surrounding gas, according to: 
\begin{equation} 
\label{eq-Edot-Feed} 
\dot{E}_{\rm feed} = \epsilon_f \epsilon_r \dot{M}_{\rm BH} c^2. 
\end{equation} 
Here $\epsilon_r$ is the radiative efficiency, 
and $\epsilon_f$ is the feedback efficiency. 
We adopt $\epsilon_r = 0.1$, which assumes radiatively efficient accretion 
onto a Schwarzschild BH \citep{Shakura73}. 

Kinetic BH feedback is included \citep[][]{Barai14, Barai16}, where the 
neighboring gas is pushed outward with a velocity $v_w$ and mass outflow rate $\dot{M}_w$. 
Using the conservation of energy, 
$\frac{1}{2} \dot{M}_w v_w^2 = \dot{E}_{\rm feed}$, 
and Eq.~(\ref{eq-Edot-Feed}), 
the BH kinetic outflow rate can be written as, 
\begin{equation} 
\label{eq-MdotW-EDW} 
\dot{M}_w = 2 \epsilon_f \epsilon_r \dot{M}_{\rm BH} \frac{c^2}{v_w^2} . 
\end{equation} 
We use the values: $\epsilon_f = 0.05$, and $v_w = 5000$ km/s. 

The BH kinetic feedback energy is distributed to the gas within a distance
$h_{\rm BH}$ from the BH. 
A bi-cone volume is defined around the BH, 
of slant height $h_{\rm BH}$ and half-opening angle $45^{\circ}$. 
The cone-axis direction is randomly assigned during a BH seeding, 
and remains fixed henceforth for each BH. 
The total gas mass within the bi-cone $M_{\rm gas}^{\rm vicinity}$ is computed. 
A probability is calculated for the $i$'th gas particle inside the bi-cone, 
in a timestep $\Delta t$: 
\begin{equation} 
\label{eq-probKick} 
p_i = \frac{\dot{M}_w \Delta t} {M_{\rm gas}^{\rm vicinity}} , 
\end{equation} 
where $\dot{M}_w$ is the mass outflow rate obtained from Eq.~(\ref{eq-MdotW-EDW}). 
A random number $x_i$ is drawn uniformly distributed in the interval $[0, 1]$. 
If $x_i < p_i$, then the gas particle is imparted a velocity boost by AGN wind, such that: 
\begin{equation} 
\label{eq-vNew} 
\vec{v}_{\rm new} = \vec{v}_{\rm old} + v_w \hat{n} . 
\end{equation} 
The AGN wind direction $\hat{n}$ is considered radially outward from the BH. 

We do not incorporate any scheme for BH {\it pinning}. 
Each BH is not repositioned at each time-step to the location of the 
minimum gravitational potential of its host galaxy. 

We consider that when galaxies merge during hierarchical structure assembly, 
their hosted central BHs merge as well.
In the numerical algorithm, two BH particles are allowed to merge 
to form a single BH, when the distance between them is smaller than the 
smoothing length of either one and their relative velocity is 
below the local sound speed \citep[e.g.,][]{Sijacki07, DiMatteo12}. 
The final merged BH has a mass equal to the sum of the BH masses, 
and a velocity along the center of mass of the initial two merging BHs. 
To impart kinetic feedback energy, the merged BH retains the 
bi-cone axis direction of the more massive initial BH. 

\subsection{Simulations} 
\label{sec-simulation} 

We perform cosmological hydrodynamical simulations of $(50 ~ {\rm Mpc})^3$ comoving volumes. 
Our standard resolution is eight simulations using 
$256^3$ dark matter and $256^3$ gas particles in the initial condition. 
The dark matter particle mass is $m_{\rm DM} = 2.47 \times 10^{8} M_{\odot}$, 
and the gas particle mass is $m_{\rm gas} = 4.61 \times 10^{7} M_{\odot}$. 
The gravitational softening length is set as $L_{\rm soft} = 3.3$ kpc comoving. 
The {\sc MUSIC}\footnote{MUSIC - Multi-scale Initial Conditions for Cosmological 
Simulations: https://bitbucket.org/ohahn/music} software \citep{Hahn11} 
is used to generate the initial condition at $z = 100$. 
The concordance flat $\Lambda$CDM cosmological model is used: 
$\Omega_{M, 0} = 0.3089, \Omega_{\Lambda, 0} = 0.6911, \Omega_{B, 0} = 0.0486, 
H_{0} = 67.74$ km s$^{-1}$ Mpc$^{-1}$ \citep[][results XIII]{Planck15}. 
Starting from $z = 100$, the boxes are subsequently evolved up to $z = 0$, 
with periodic boundary conditions. 

% In addition, we execute one high-resolution simulation with $512^3$ dark matter and $512^3$ gas particles. 
% (dark matter particle mass for the high-resolution case $3.1 \times 10^{7} M_{\odot}$)
% (gas particle mass for the high-resolution run $5.8 \times 10^{6} M_{\odot}$)
% (in the high-resolution run $1.6$ kpc comoving) 

We execute a series of simulations, with characteristics listed in Table~\ref{Table-Sims}. 
All the runs incorporate metal cooling, chemical enrichment, and SF. 
The first run {\it SFonly} has no SN feedback neither BHs, 
SN feedback is additionally included in the second run, 
while the remaining runs additionally include BHs. 

\begin{itemize} 

\item {\it SN} : no BH present. This is a control simulation in which 
only cooling, star-formation, chemical evolution, and stellar/SN feedback are implemented. 
\item {\it BHsXeYvZ} : with BH accretion and feedback. 
The symbols $X$, $Y$ and $Z$ indicate the parameters being varied between the simulations 
in the following way, $M_{\rm BHseed} = 10^{X} M_{\odot}$, BH feedback efficiency $\epsilon_f = Y$, 
and BH kinetic feedback kick velocity $v_w = Z \times 1000$ km/s. 

\end{itemize} 

Halos are identified by executing a {\it Friends-of-Friends} (FOF) 
group finder on-the-fly within our simulations. 
Galaxies are tracked using the subhalo finder {\it SubFind}, 
which associates substructures to FOF halos. 
The centre of each galaxy is considered as the location of 
the gravitational potential minimum of its subhalo. 
The halo mass $(M_{\rm halo})$ of a galaxy, and its virial radius in comoving 
coordinates $(R_{200})$, are related such that $R_{200}$ encloses a density 
$200$ times the mean comoving matter density of the Universe: 
\begin{equation} 
\label{eq-Mhalo} 
M_{\rm halo} = \frac{4 \pi}{3} R_{200}^3 \left(200 \rho_{\rm crit} \Omega_{M,0}\right) , 
\end{equation} 
where $\rho_{\rm crit} = 3 H_0^2 / (8 \pi G)$ is the present critical density. 
The galaxy stellar mass is considered as the mass of all star particles inside 
the subhalos obtained by the subhalo finder {\it SubFind}.

%%%%%%%%%%%%%%%%%%%%%%%%%%%%%%%%%%%%%%%%%%%%%%%%%%%%%%%%%%%%%%%%%%%%%%%% 
% FIGURE 1 
\begin{figure*} 
\centering 
\includegraphics[width = 1.0 \linewidth]{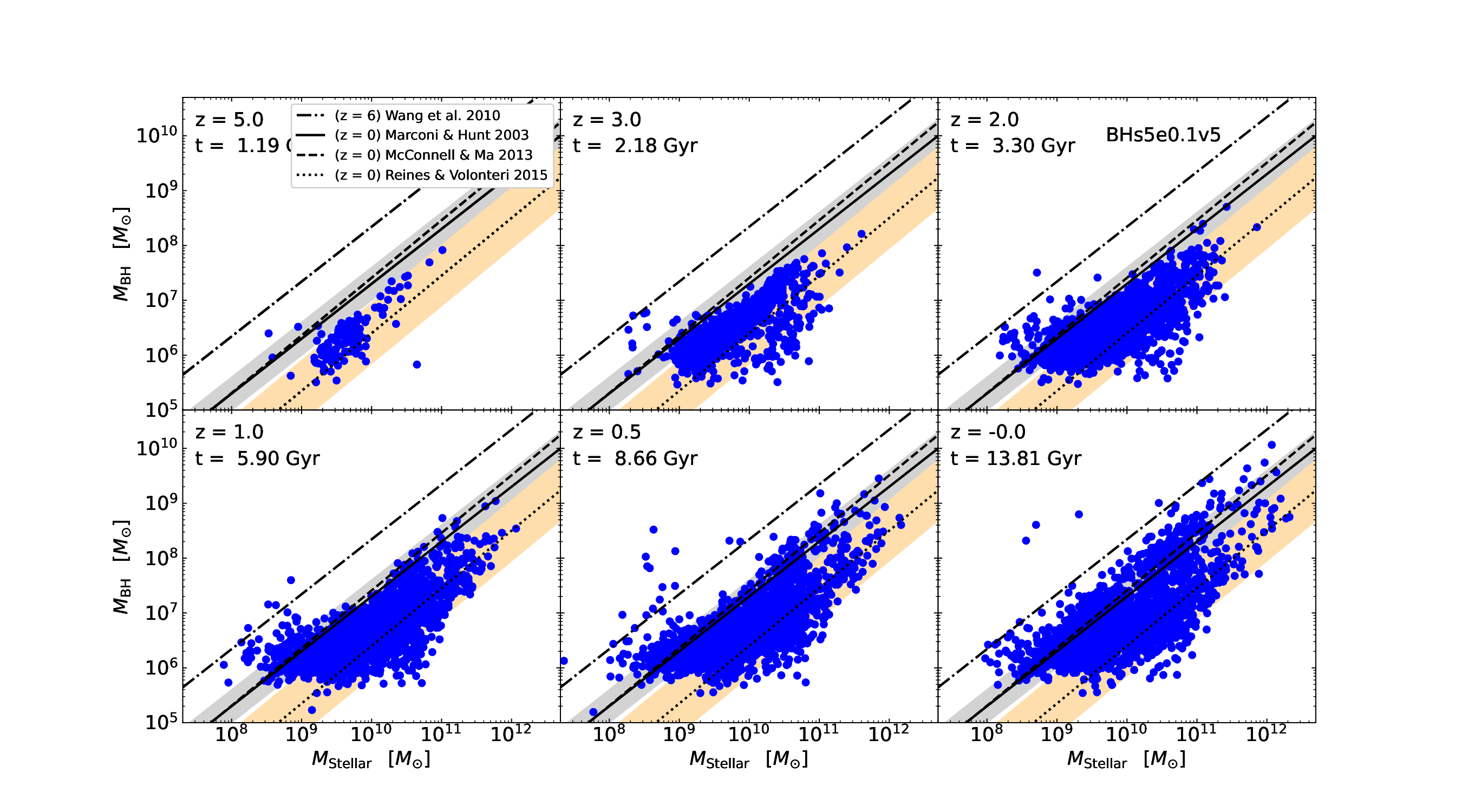} 
\vspace{-1.2cm} 
\caption{ 
Cosmic build-up of the BH mass versus stellar mass of our simulated galaxies 
in run {\it BHs5e0.1v5}, each panel indicating one redshift $z = 5, 3, 2, 1, 0.5, 0$. 
The blue circles are galaxies from our simulation. 
The black lines indicate the observed BH mass versus galaxy stellar mass relation for: 
local galaxies showing the correlation with the bulge mass as the solid \citep{Marconi03}, 
dashed \citep{McConnell13}, and dotted \citep{Reines15} lines; 
as well as $z \sim 6$ quasars \citep{Wang10} as the dash-dotted line. 
} 
\label{fig-Mass-BH-vs-Stellar-Evol-1run}
\end{figure*}
%%%%%%%%%%%%%%%%%%%%%%%%%%%%%%%%%%%%%%%%%%%%%%%%%%%%%%%%%%%%%%%%%%%%%%%% 

%%%%%%%%%%%%%%%%%%%%%%%%%%%%%%%%%%%%%%%%%%%%%%%%%%%%%%%%%%%%%%%%%%%%%%%% 
% FIGURE 2 
\begin{figure*} 
\centering 
\includegraphics[width = 0.81 \linewidth]{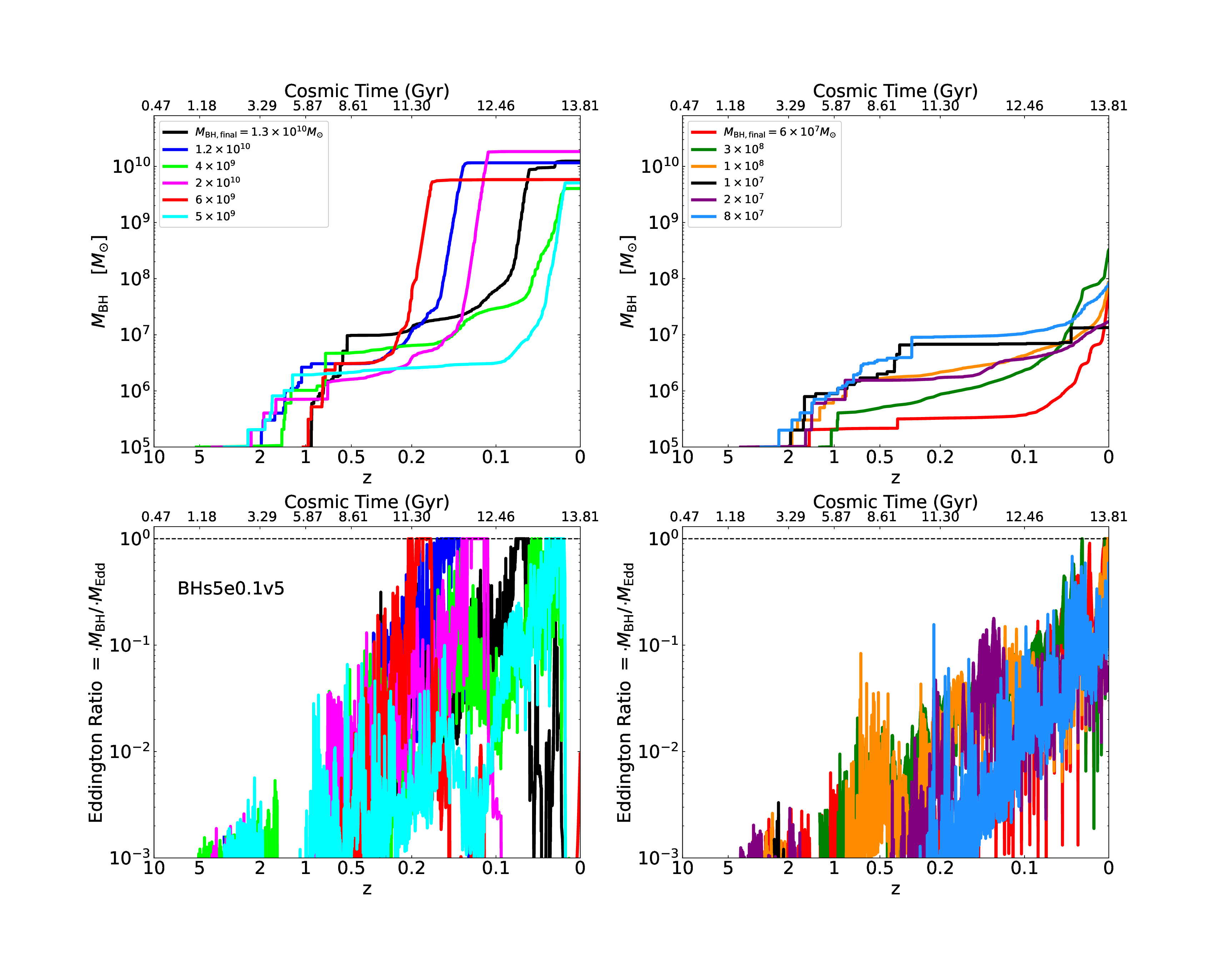} 
\vspace{-1.2cm} 
\caption{ 
Growth with redshift of BH mass (top row) and Eddington accretion ratio (bottom row) 
of some of the massive BHs in our standard simulation {\it BHs5e0.1v5}. 
The left column indicate six most-massive BHs with final mass $M_{\rm BH (z=0)} > 10^{9} M_{\odot}$, 
and the right column shows six intermediate-mass BHs of $M_{\rm BH (z=0)} = (10^{7} - 10^{8}) M_{\odot}$. 
}
\label{fig-BH-Mass-EddRatio-vs-z}
\end{figure*}
%%%%%%%%%%%%%%%%%%%%%%%%%%%%%%%%%%%%%%%%%%%%%%%%%%%%%%%%%%%%%%%%%%%%%%%%  

\section{Results and Discussion} 
\label{sec-results} 

% Black Hole Mass vs. Galaxy Stellar Mass Relation Build-Up with Cosmic Time 
\subsection{$M_{\rm BH}$ vs. $M_{\star}$ Relation Build-Up with Cosmic Time} 

The cosmic build-up of the BH mass $M_{\rm BH}$ versus stellar mass $M_{\star}$ of our simulated galaxies 
is shown in Fig.~\ref{fig-Mass-BH-vs-Stellar-Evol-1run}. 
Six epochs are plotted of our standard run {\it BHs5e0.1v5} 
(which gives the best-fit to the observational \citet{Marconi03} $z = 0$ correlation), 
one each at $z = 5, 3, 2, 1, 0.5, 0$. 
Observational data is overplotted as the black lines
indicating the BH mass versus stellar bulge mass relationships at different epochs.
Local galaxies ($z = 0$) are represented by the solid line:
$M_{\rm BH} / M_{\star} = 0.002$ \citep{Marconi03},
as well as from \citet{Reines15} as the dotted line,
and from \citet{McConnell13} as the dashed line. 
The ratio is observed to be steeper at high-$z$. 
Bright $z \sim 6$ quasars, observed in the far-IR and CO, lie along the dash-dotted line: 
median $M_{\rm BH} / M_{\star} = 0.022$ \citep{Wang10}. 

We find the slope of the $[M_{\rm BH} - M_{\star}]$ correlation 
remains almost constant in our simulations, and there is an intercept evolution with cosmic time. 
Noting that the observational relation by \citet{Reines15} lies somewhat lower than that by 
\citet{Marconi03} and \citet{McConnell13}; 
consequently our simulated galaxies are found to follow these 
different relations at distinct redshifts. 
At earlier epochs $z = 5, 3$ the BHs follow more closely the \citet{Reines15} correlation, 
at $z = 2 - 1$ the BHs are in between the two relations, 
while at current epochs $z = 0$ they lie more on the \citet{Marconi03} and \citet{McConnell13} correlations. 

%%%%%%%%%%%%%%%%%%%%%%%%%%%%%%%%%%%%%%%%%%%%%%%%%%%%%%%%%%%%%%%%%%%%%%%% 
% FIGURE 3 --- NEW FIGURE INSERTED AFTER REFEREE REPORT  
\begin{figure*} 
\centering 
\includegraphics[width = 0.96 \linewidth]{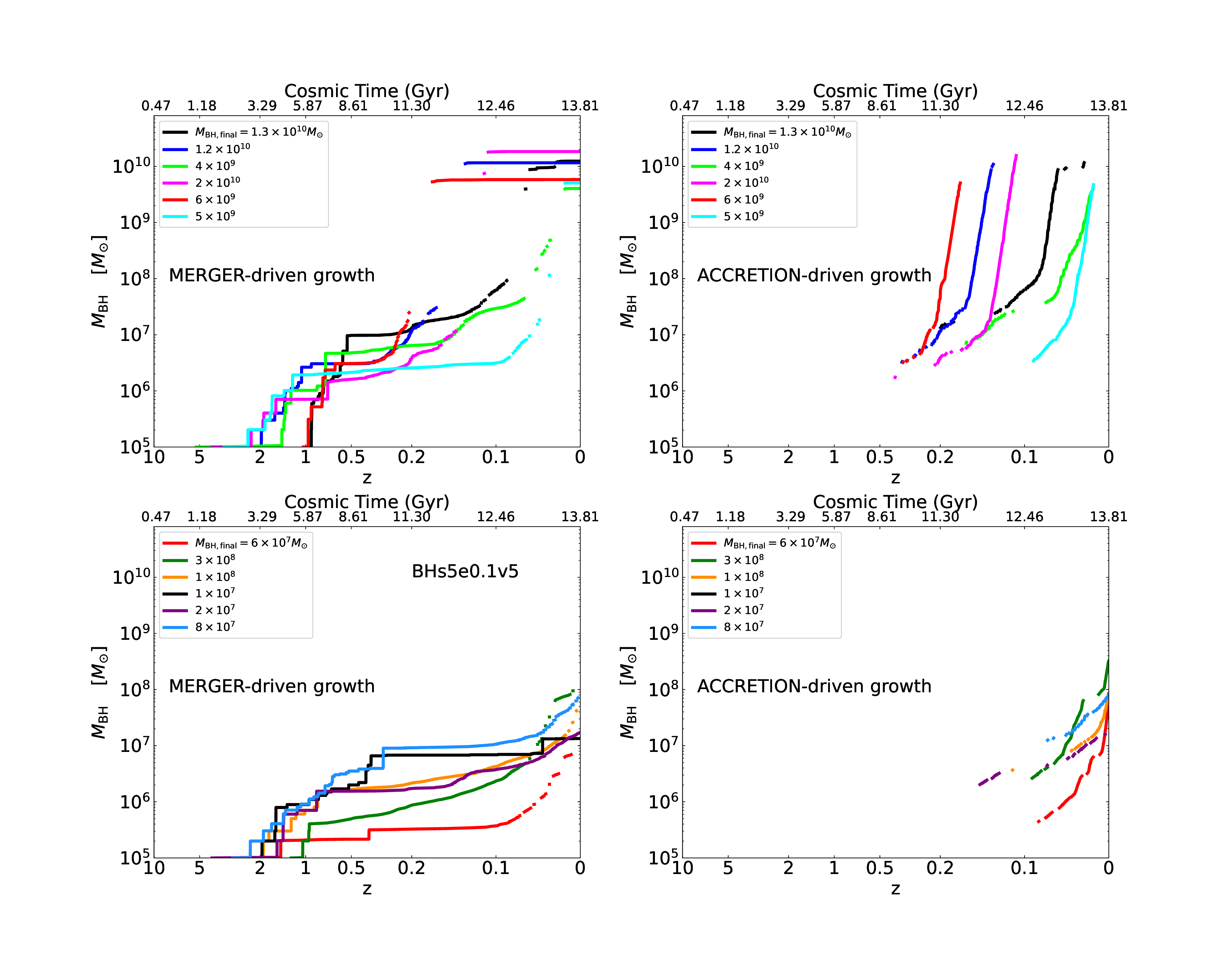} 
\vspace{-1.2cm} 
\caption{ 
Growth of BH mass due to Mergers or Accretion of the same BHs as plotted in Fig.~\ref{fig-BH-Mass-EddRatio-vs-z}: 
six most-massive BHs (top row) and six intermediate-mass BHs (bottom row) from our standard simulation {\it BHs5e0.1v5}. 
The left column shows the Merger-Dominated mass growth regime and 
the right column presents the Accretion-Dominated mass growth regime; 
the two regimes distinguished with a limiting Eddington accretion ratio $= 0.1$ as described in \S\ref{sec-res-MD-vs -AD}. 
}
\label{fig-BH-Mass-Merger-vs-Accretion}
\end{figure*}
%%%%%%%%%%%%%%%%%%%%%%%%%%%%%%%%%%%%%%%%%%%%%%%%%%%%%%%%%%%%%%%%%%%%%%%%  

\subsection{Mass Growth of the Supermassive Black Holes: {\bf Merger-driven versus  Accretion-driven Regimes}} 
\label{sec-res-MD-vs -AD} 

In our simulations, the first BHs are seeded at $z \sim 9$. 
Generally in such cosmological simulations, the redshift of the seeding of BHs 
is a quantity dependent on periodic boxsize and numerical resolution. 
For our standard resolution runs, $z = 8.5 - 9$ is 
the first epoch when a massive halo reach $M_{\rm halo} = 10^{10} M_{\odot}$; 
henceforth a BH of $10^{5} M_{\odot}$ is seeded at its center. 
More BHs are seeded at later epochs following the prescription described in the methods. 

We find that none of the first seeds (those BHs that are seeded at $z \sim 9$) 
grow to become one of the most-massive BHs at $z = 0$. 
The BHs which become most-massive are actually seeded at later epochs $z \sim 2 - 5$. 
This is because of the different BH growth modes, as described next. 

The growth with redshift of some of the massive BHs in our {\it BHs5e0.1v5} run 
is plotted in Fig.~\ref{fig-BH-Mass-EddRatio-vs-z}: BH mass in the top row, 
and Eddington accretion ratio $= \dot{M}_{\rm BH} / \dot{M}_{\rm Edd}$ in the bottom row. 
Each BH starts from an initial seed of $M_{\rm BH} = 10^5 M_{\odot}$. 
The subsequent mass growth is due to merger with other BHs (revealed as vertical rises in $M_{\rm BH}$),
and gas accretion (visualized as the positive-sloped regions of the $M_{\rm BH}$ versus $z$ curve). 

We quantify the driving mechanism of the mass growth of the BHs using a limiting Eddington accretion ratio of $0.1$. 
If $\dot{M}_{\rm BH} / \dot{M}_{\rm Edd} \geq 0.1$ then it is considered as an Accretion-driven mass growth regime; 
while $\dot{M}_{\rm BH} / \dot{M}_{\rm Edd} < 0.1$ is considered as a Merger-Driven growth regime. 
The results of these two growth regimes are plotted in Fig.~\ref{fig-BH-Mass-Merger-vs-Accretion}. 
The left column shows the  mass growth due to mergers with other BHs and 
the right column presents the mass growth due to gas accretion, of six most-massive BHs (top row) 
and six intermediate-mass BHs (bottom row) from our standard simulation {\it BHs5e0.1v5} 
(the same BHs as plotted in Fig.~\ref{fig-BH-Mass-EddRatio-vs-z}). 

We find that the initial growth from $10^5 M_{\odot}$ seeds to $\sim 10^7 M_{\odot}$ 
occurs predominantly via BH mergers (left two panels of Fig.~\ref{fig-BH-Mass-Merger-vs-Accretion}). 
For the BHs that become supermassive with mass $\geq 10^9 M_{\odot}$, 
their final growth occurs over a period of $600 - 700$ Myr dominated by efficient gas accretion 
(top-right panel of Fig.~\ref{fig-BH-Mass-Merger-vs-Accretion}). 
This period corresponds to an Eddington-limited accretion where Eddington ratio $= 1$ 
(bottom-left panel of Fig.~\ref{fig-BH-Mass-EddRatio-vs-z}). 
During this period, the BH mass increases by a factor $10^2 - 10^3$; which grows the BHs 
from $10^7 M_{\odot}$ to $(10^9 - 10^{10}) M_{\odot}$ in a short time of $600 - 700$ Myr. 
After this point, the SMBHs have a flat mass evolution with redshift 
(top-left panel of Fig.~\ref{fig-BH-Mass-Merger-vs-Accretion}), i.e. they stop growing up to $z = 0$. 
We speculate this final growth halt happens because of gas depletion from galaxy centers. 

The cosmic epoch of this rapid efficient gas accretion is also seen to vary depending on the SMBH, 
as can be seen in the top-right panel of Fig.~\ref{fig-BH-Mass-Merger-vs-Accretion}. 
E.g., the red curve grows from $M_{\rm BH} = 7 \times 10^{6} M_{\odot}$ at $z = 0.4$ 
to $M_{\rm BH} = 6 \times 10^{9} M_{\odot}$ at $z \sim 0.3$ and has a passive evolution thereafter. 
While the cyan curve grows from $M_{\rm BH} = 4 \times 10^{6} M_{\odot}$ at $z = 0.1$ 
to $M_{\rm BH} = 5 \times 10^{9} M_{\odot}$ at $z = 0.05$. 
The consistent feature among all the BHs is that the final growth inevitably 
happens within a period of $600 - 700$ Myr (as described in the previous paragraph). 

We argue the physical reason behind such BH growth behavior is the influence of 
gas supply to the host galaxy center, and subsequent efficient gas accretion onto the BH, 
leading to gas depletion within $700$ Myr. 

The right column of Fig.~\ref{fig-BH-Mass-EddRatio-vs-z} and 
the bottom row of Fig.~\ref{fig-BH-Mass-Merger-vs-Accretion} presents six intermediate-mass BHs: 
those which reach $M_{\rm BH (z=0)} = (10^{7} - 10^{8}) M_{\odot}$ at the present epoch. 
We find that their growth have happened primarily via BH mergers 
(indicated by vertical rises of $M_{\rm BH}$). 
Their accretion ratio remains sub-Eddington. 
At $z < 0.05$, some of them (blue, green, red, orange curves) reach Eddington ratio $= 1$ 
for $100 - 200$ Myr, but soon becomes sub-Eddington because of gas depletion. 

\subsubsection{Implications for the Little Red Dots} 

We can apply our numerical simulation results of SMBH growth 
to the $z > 4$ little-red-dots recently observed by JWST \citep[e.g.,][]{Labbe2023, Killi2024, Matthee2024}: 
sources with a compact morphology, red optical color having a V-shaped spectral energy distribution, 
and spectroscopic broad Balmer emission lines. 
The observed emission of the little-red-dots likely consist of an AGN component \citep[e.g.,][]{Greene2024, Xiao2025} 
with BH mass in the range $M_{\rm BH} = 10^{6} - 10^{8} M_{\odot}$, 
and/or a compact starburst \citep[e.g.,][]{PerezGonzalez2024, Williams2024}. 

With the AGN source interpretation, the little-red-dots represent low-luminosity AGN \citep[e.g.,][]{Ma2025}; 
which are often obscured by dust \citep[e.g.,][]{Fujimoto2022, Akins2023} in the early Universe. 
These high-$z$ faint AGN comprise of a surprisingly abundant population at $z > 5$; 
with a volume number density $\sim 100$ times the extrapolated quasar UV luminosity function 
\citep[e.g.,][]{Matthee2024, Lin2024}. 
As a fraction of the galaxy population, \citet{Harikane2023} found that 
$\sim 5\%$ of galaxies at $z = 4-7$ are type-1 AGN with broad lines, 
which is thus a higher fraction than $z \sim 0$ galaxies with similar luminosities. 

When the host galaxy stellar masses of the little-red-dots are inferred from observations, 
and compared to the local $[M_{\rm BH} - M_{\star}]$ correlation, 
it implies a fast black hole growth at early epochs \citep{Harikane2023}. 
Interpreting with our numerical simulation results on the growth modes of the BHs presented \S\ref{sec-res-MD-vs -AD}: 
the high-$z$ BH growth observed in studies like \citep{Harikane2023} must have been dominated by efficient gas accretion. 

Our implications of the little-red-dots assumes the AGN source interpretation. 
These JWST-detected early SMBHs at $z \sim 4-12$ comprise of some BHs 
which are overmassive \citep[e.g.,][]{Goulding2023, Wu2025} in their host galaxy, as well as some 
normal-mass SMBHs \citep[e.g.,][]{Li2025, Geris2025} with respect to the local $M_{\rm BH} - M_{\star}$ correlation. 
Applying our results to these high-$z$ AGN in Little Red Dots, we assert that: 
the normal-mass SMBHs had predominantly undergone BH merger driven evolution, 
whereas the overmassive BHs underwent periods of Eddington-limited 
or super-Eddington bursts of efficient gas accretion. 
The high nuclear gas density in the galaxies forming in the early Universe 
lead to effective gas accretion onto BHs turning them overmassive. 
These trends also imply that the $[M_{\rm BH} - M_{\star}]$ relation 
becomes more defined (i.e. with a smaller scatter) at lower redshifts. 

\subsection{Cosmic Evolution along the $[M_{\rm BH} - M_{\star}]$ diagram} 

\begin{figure*} 
\centering 
\includegraphics[width = 0.75 \linewidth]{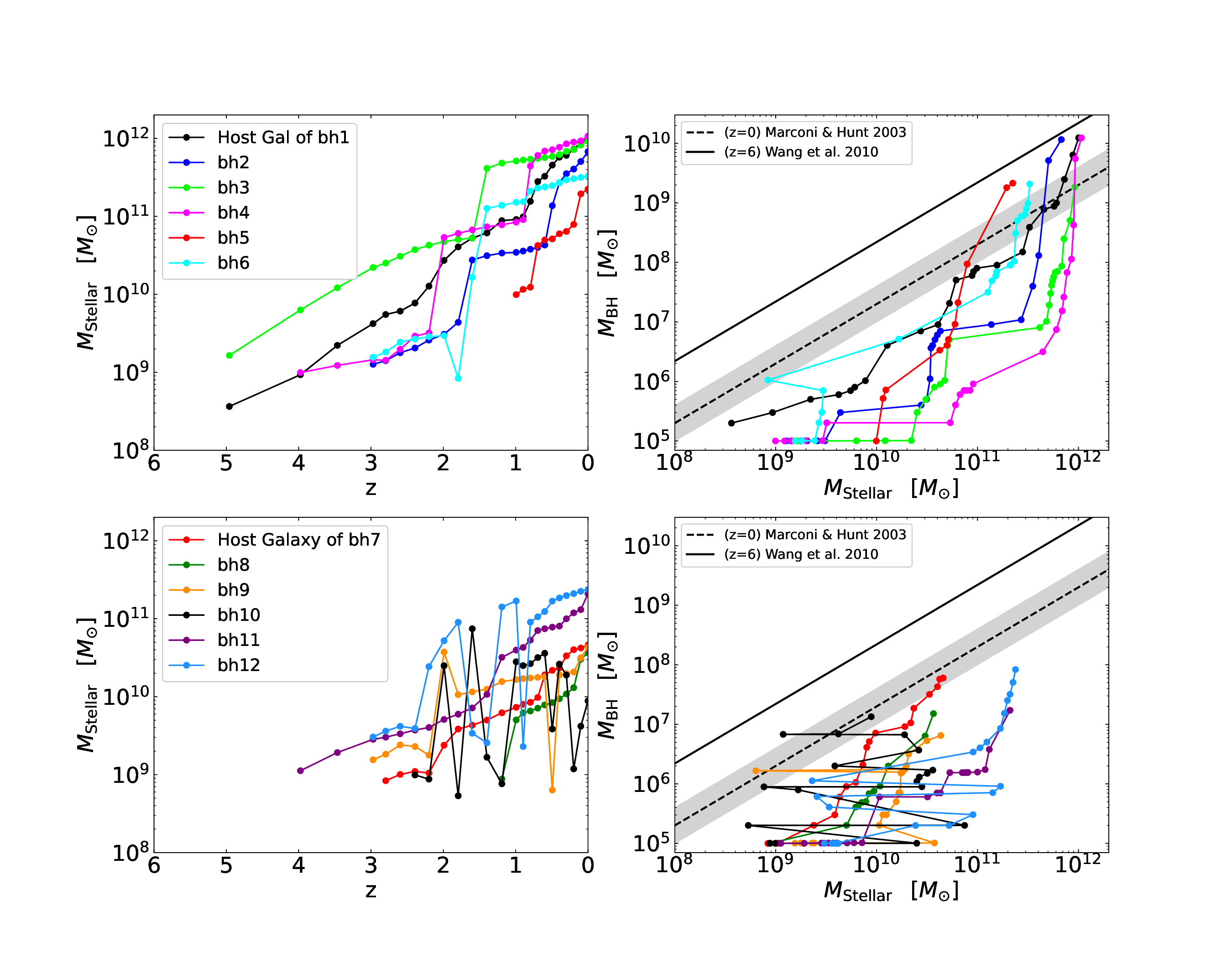} 
\vspace{-1.0cm} 
\caption{ 
Left column: Build up with redshift of the stellar mass of the host galaxy 
of the six most-massive BHs (top row) and six intermediate-mass BHs (bottom row); 
same BHs which were plotted in Fig.~\ref{fig-BH-Mass-EddRatio-vs-z}, distinguished by the plotting colours. 
Right column: Redshift track or the evolution with cosmic time 
of the BH mass versus host galaxy stellar mass of the black holes. 
The black lines indicate the observed BH mass versus galaxy stellar mass relation of: 
local galaxies \citep{Marconi03} as the dashed line, and $z=6$ quasars \citep{Wang10} as the solid line. 
} 
\label{fig-Redshift-Track-Mass-Stellar-and-BH}
\end{figure*}

The right panels of Fig.~\ref{fig-Redshift-Track-Mass-Stellar-and-BH} show 
the evolution track with cosmic time of the BH mass versus host galaxy stellar mass of the 
six most-massive BHs (top row) and six intermediate-mass BHs (bottom row); 
these are same BHs which were plotted in Fig.~\ref{fig-BH-Mass-EddRatio-vs-z}. 
The corresponding redshift evolution of the host galaxy stellar mass of each BH
is shown in the left panels of Fig.~\ref{fig-Redshift-Track-Mass-Stellar-and-BH}. 
We can visualize (top-left panel) that each SMBH mostly stay in the same host galaxy which builds up with time, 
and might undergo merger to form a larger galaxy, when the stellar mass increases sharply. 
There can be occasions when the BH migrates to a less-massive host galaxy, for a 
short period of time, which happens with $bh6$ (the cyan curve) at $z \sim 1.8$. 

The temporary migration of a BH from a more-massive host galaxy to a less-massive galaxy 
happens much more frequently with the IMBHs (bottom-left panel); $bh8$, $bh9$, $bh10$, and $bh12$, 
demonstrate such migration signatures when the stellar mass decreases abruptly for some time. 
Such host migration happens more frequently for the IMBHs as compared to the SMBHs 
because of a greater influence of dynamical forces. 

In the right panels of Fig.~\ref{fig-Redshift-Track-Mass-Stellar-and-BH}, 
each track starts at a point in the bottom-left which corresponds to an epoch 
when the BH was seeded $z \sim 3 - 5$, and evolve up to the top-right point $z = 0$. 
After seeding, the BHs start below the observed $[M_{\rm BH} - M_{\star}]$ correlation, 
and they slowly evolve to the $z = 0$ correlation first, 
and some BHs eventually toward the $z = 6$ correlation. 
The SMBHs (top-right panel) has reached the observed correlation at $z = 0$. 
While the IMBHs (bottom-right panel) still lies below the observed correlation. 
% keeping in mind of the large observational scatter.  

%%%%%%%%%%%%%%%%%%%%%%%%%%%%%%%%%%%%%%%%%%%%%%%%%%%%%%%%%%%%%%%%%%%%%%%% 
% FIGURE 4 
%%%%%%%%%%%%%%%%%%%%%%%%%%%%%%%%%%%%%%%%%%%%%%%%%%%%%%%%%%%%%%%%%%%%%%%% 

%%%%%%%%%%%%%%%%%%%%%%%%%%%%%%%%%%%%%%%%%%%%%%%%%%%%%%%%%%%%%%%%%%%%%%%% 
% FIGURE 5 
\begin{figure*}
\centering
\includegraphics[width = 0.95 \linewidth]{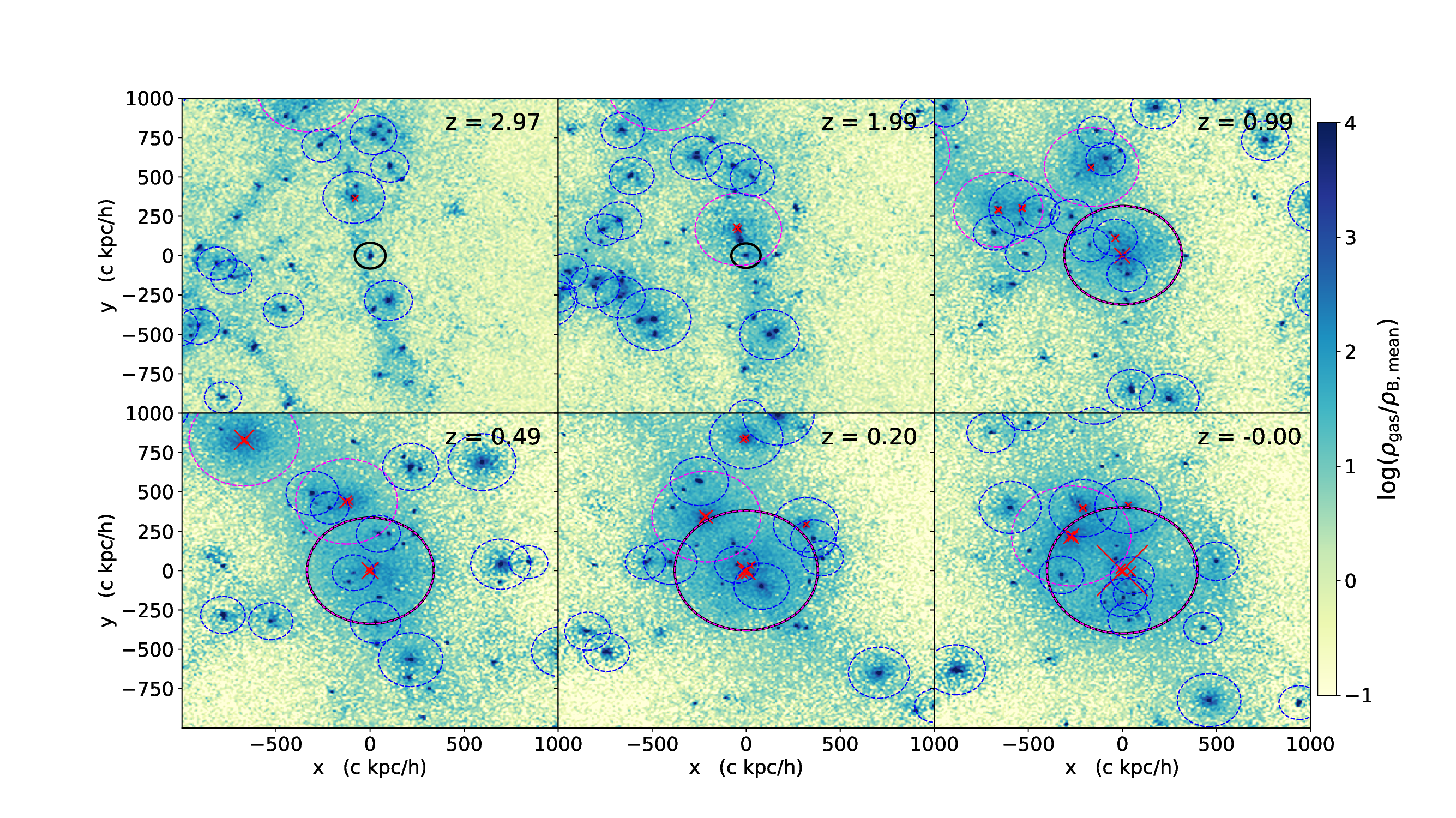} 
\vspace{-0.5cm} 
\caption{ 
Gas overdensity in our standard simulation {\it BHs5e0.1v5} at six epochs $z = 3, 2, 1, 0.5, 0.2, 0$. 
The BH which would become most-massive at $z=0$ is back tracked, 
and each panel shows a projected $(2000 h^{-1}$ kpc$)^3$ comoving volume around the location of this BH. 
The red cross symbols indicate the positions of all the BHs inside the plotted region, 
with the symbol size proportional to BH mass. 
The black circle is the virial radius $R_{\rm 200}$ of the host galaxy of the back-tracked BH, 
the red circles depict the $R_{\rm 200}$ of the galaxies with the mass range $M_{\rm halo} > 10^{12} M_{\odot}$, 
while the blue circles show the $R_{\rm 200}$ of $(10^{11} < M_{\rm halo} < 10^{12}) M_{\odot}$ galaxies. 
} 
\label{fig-GasOden-BHpos} 
\end{figure*} 
%%%%%%%%%%%%%%%%%%%%%%%%%%%%%%%%%%%%%%%%%%%%%%%%%%%%%%%%%%%%%%%%%%%%%%%% 

%%%%%%%%%%%%%%%%%%%%%%%%%%%%%%%%%%%%%%%%%%%%%%%%%%%%%%%%%%%%%%%%%%%%%%%% 
% FIGURE 6 
\begin{figure*} 
\hspace{-1.5cm} 
\includegraphics[width = 1.2 \linewidth]{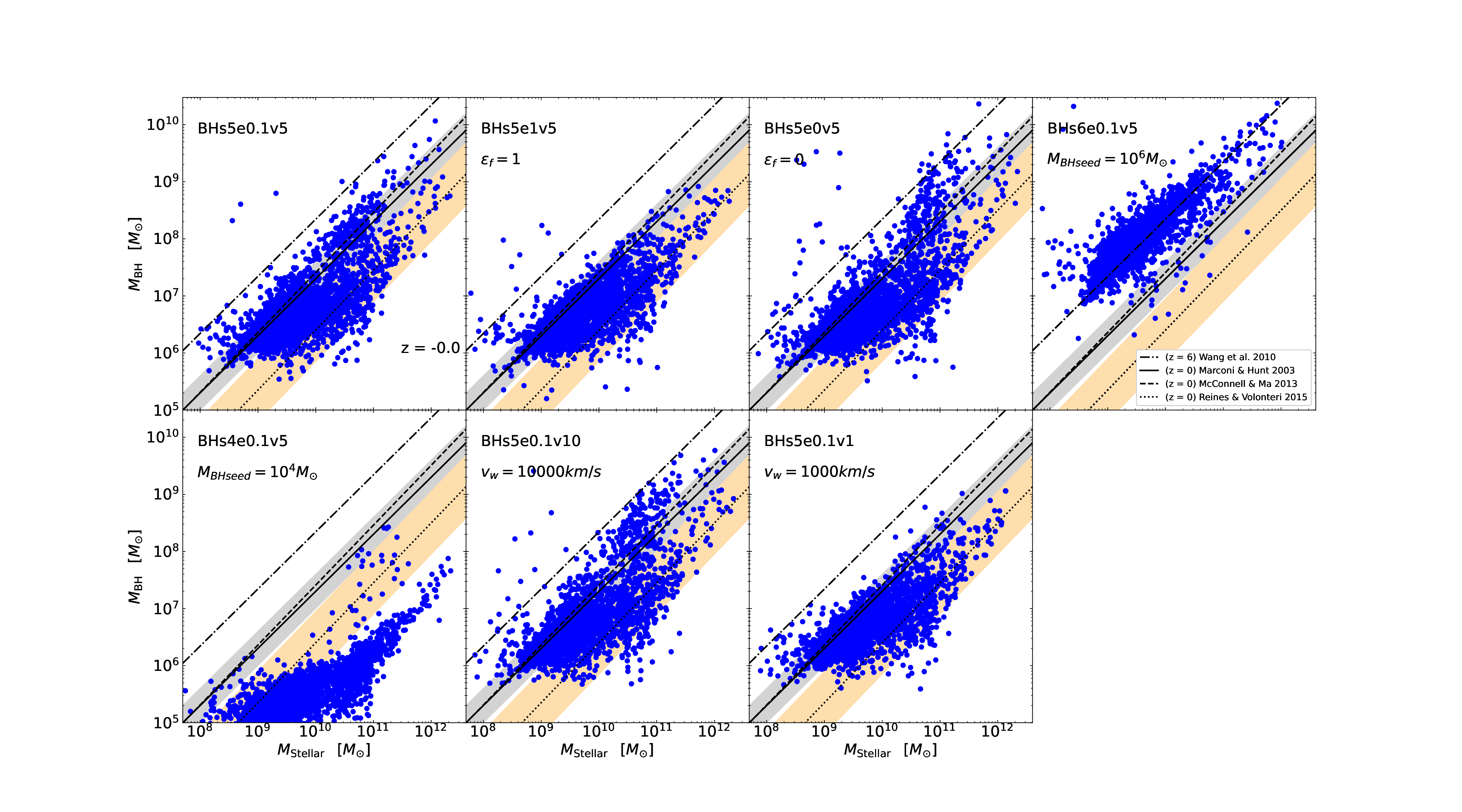} 
\vspace{-1.2cm} 
\caption{
BH mass versus stellar mass of the galaxies obtained in our simulations, 
at the current epoch $z = 0$, each panel indicating one distinct run. 
The black lines indicate the observed BH mass versus galaxy stellar mass relation for: 
local galaxies showing the correlation with the bulge mass as the solid \citep{Marconi03}, 
dashed \citep{McConnell13}, and dotted \citep{Reines15} lines; 
as well as $z \sim 6$ quasars \citep{Wang10} as the dash-dotted line. 
}
\label{fig-Mass-BH-vs-Stellar-z0-6runs}
\end{figure*}
%%%%%%%%%%%%%%%%%%%%%%%%%%%%%%%%%%%%%%%%%%%%%%%%%%%%%%%%%%%%%%%%%%%%%%%% 

\subsection{Environment of the BHs} 

The evolution of gas environment around an assembling massive BH is plotted in Fig.~\ref{fig-GasOden-BHpos}. 
It displays gas overdensity (i.e., the ratio between the gas density 
and the cosmological mean baryon density of the Universe) 
in our simulation {\it BHs5e0.1v5}, at six epochs $z = 3, 2, 1, 0.5, 0.2, 0$. 
The BH which would become most-massive in this run at $z=0$ is back tracked, and each panel 
shows a projected $(2000 h^{-1}$ kpc$)^3$ comoving volume around the location of this BH. 
The black circle is the virial radius $R_{\rm 200}$ (defined in Eq.~\ref{eq-Mhalo}) 
of the host galaxy of the back-tracked BH, 
the red circles depict the $R_{\rm 200}$ of the galaxies with the mass range $M_{\rm halo} > 10^{12} M_{\odot}$, 
while the blue circles show the $R_{\rm 200}$ of $(10^{11} < M_{\rm halo} < 10^{12}) M_{\odot}$ galaxies. 
The spatial locations of all the BHs inside the plotted region can be visualized with the red cross symbols, 
with the symbol size proportional to BH mass. 

In the top panels of Fig.~\ref{fig-GasOden-BHpos} we can see the cosmological 
large-scale-structure Mpc-scale filaments, consisting of dense (blue and dark-blue regions) gas. 
The galaxies (blue and red circles) lie along the filaments, 
or at the high-density intersections of the filaments, 
and finally in the galaxy cluster region which has formed at $z = 0$. 
The clustering of halos are visible in all the panels: 
at $z = 3, 2, 1$ the halos are forming along cosmological large-scale-structure filaments. 
While at $z = 0.5, 0.2$ the halos have formed a overdense proto-cluster region at the center 
of the plotted volume, which evolves to a dense galaxy cluster region at $z = 0$. 
Thus the most-massive BH in this simulation of 
$M_{\rm BH (z=0)} = 2 \times 10^{10} M_{\odot}$ lies at the center of a galaxy cluster.

% Influence of Feedback Model Parameters on the Black Hole Mass - Galaxy Stellar Mass Correlation 
\subsection{Influence of AGN Feedback Model Parameters} 

The BH mass versus galaxy stellar mass correlation obtained 
in 7 of our simulations at the current epoch $z = 0$ 
is presented in Fig.~\ref{fig-Mass-BH-vs-Stellar-z0-6runs}. 
We find a relatively large scatter in the $[M_{\rm BH} - M_{\star}]$ 
correlation of our simulated galaxies, which is comparable 
to the scatter found in observations \citep[e.g.,][]{Reines15}. 
Our standard simulation {\it BHs5e0.1v5} with $M_{\rm BHseed} = 10^{5} M_{\odot}$, 
$\epsilon_f = 0.1$ and $v_w = 5000$ km/s; 
is able to well reproduce the observational \citep{Marconi03} $z = 0$ correlation. 

Among the relevant BH subgrid model parameter variations, the largest impact is seen with the BH seed mass. 
With a higher $M_{\rm BHseed} = 10^{6} M_{\odot}$ (run {\it BHs6e0.1v5}), 
the resulting BHs are too massive and overshoot the local relation. 
On seeding with a lower seed mass of $M_{\rm BHseed} = 10^{4} M_{\odot}$ (run {\it BHs4e0.1v5}), 
the resulting BHs do not grow enough and remain below the local one. 
There is a relatively lower impact of the BH feedback efficiency and the feedback kick velocity. 
With $\epsilon_f = 1$ (run {\it BHs5e1v5}) and with $v_w = 1000$ km/s (run {\it BHs5e0.1v1}) 
the BHs lie slightly below the local $[M_{\rm BH} - M_{\star}]$ correlation. 
With $\epsilon_f = 0$ (run {\it BHs5e0v5}) and with $v_w = 10000$ km/s (run {\it BHs5e0.1v10}) 
the BHs lie slightly above the $z = 0$ local relation. 

Worthwhile to note that the simulation with $\epsilon_f = 0$ (run {\it BHs5e0v5}) is one with 
no BH feedback because the efficiency (viz. Eq.~(\ref{eq-Edot-Feed})) is set to zero, as a test case. 
From the top row, third panel of Fig.~\ref{fig-Mass-BH-vs-Stellar-z0-6runs}, 
we thus find that even with no BH feedback some kind of 
$[M_{\rm BH} - M_{\star}]$ correlation is generated in the simulations.  

% These two are also the runs where SF quenching happens the earliest 
% implying that the suppression occurs due to BH activities. 

% We find that there is a direct connection between early BH growth 
% and the quenching of SF, which is henceforth caused by resulting BH feedback. 
% In addition, our results of fast BH growth at the centers of dwarf galaxies 
% suggest that these BHs grow faster than their host galaxies in the early Universe. 

%%%%%%%%%%%%%%%%%%%%%%%%%%%%%%%%%%%%%%%%%%%%%%%%%%%%%%%%%%%%%%%%%%%%%%%% 
% FIGURE 7 
\begin{figure*} 
\centering 
\includegraphics[width = 0.8 \linewidth]{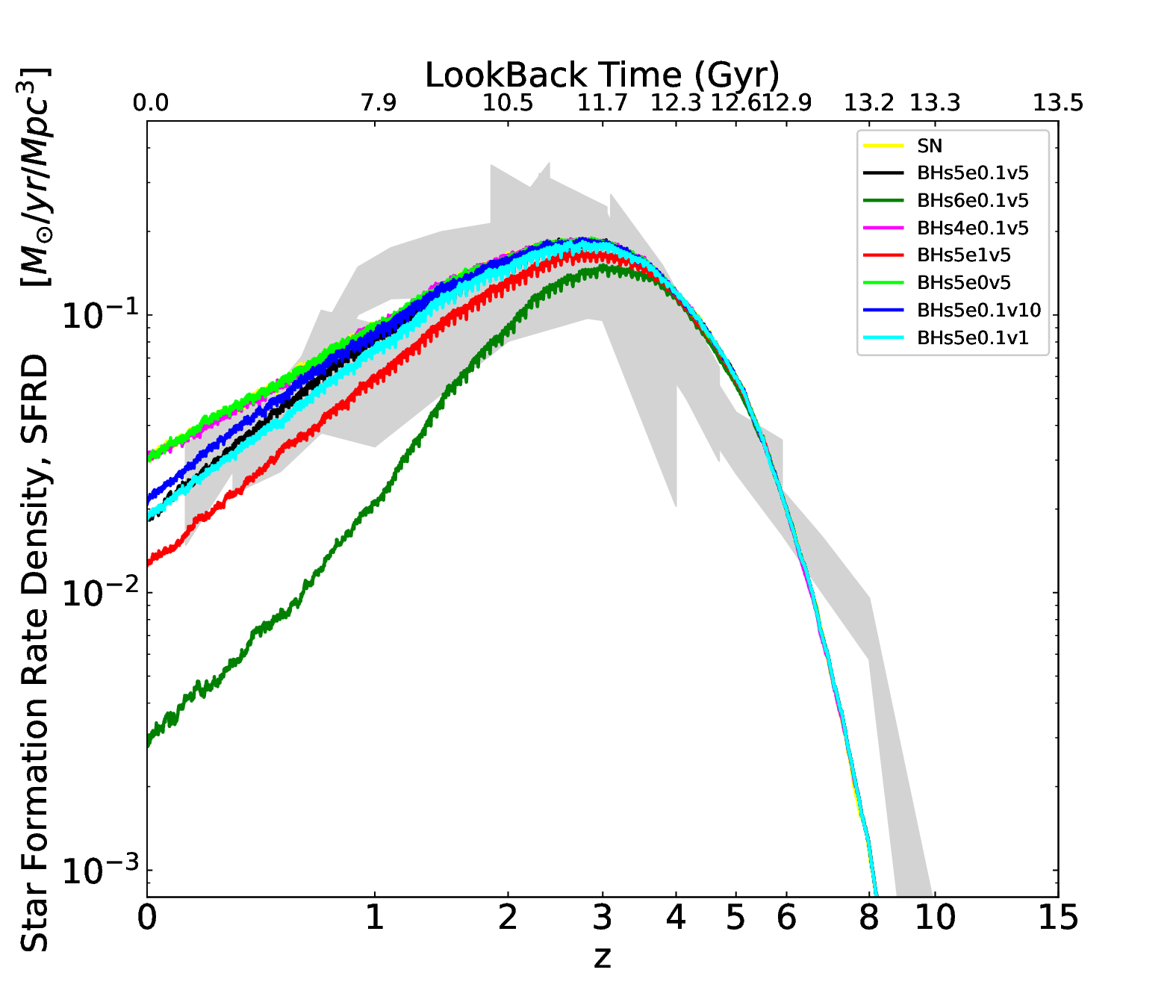} 
\vspace{-0.5cm} 
\caption{ 
Star formation rate density integrated over the whole simulation volume as a function of 
redshift, with the different simulation runs labelled by the colours and linestyles. 
The grey shaded region denotes a combination of observational SFRD data range from 
\citet{Cucciati12}, and the compilations therein originally from 
\citet{PerezGonzalez05}, \citet{Schiminovich05}, \citet{Bouwens09}, 
\citet{Reddy09}, \citet{Rodighiero10}, \citet{vanderBurg10}, \citet{Bouwens12}. 
} 
\label{fig-SFRD} 
\end{figure*} 
%%%%%%%%%%%%%%%%%%%%%%%%%%%%%%%%%%%%%%%%%%%%%%%%%%%%%%%%%%%%%%%%%%%%%%%% 

%%%%%%%%%%%%%%%%%%%%%%%%%%%%%%%%%%%%%%%%%%%%%%%%%%%%%%%%%%%%%%%%%%%%%%%% 
% FIGURE 8 
\begin{figure*} 
\centering 
\includegraphics[width = 1 \linewidth]{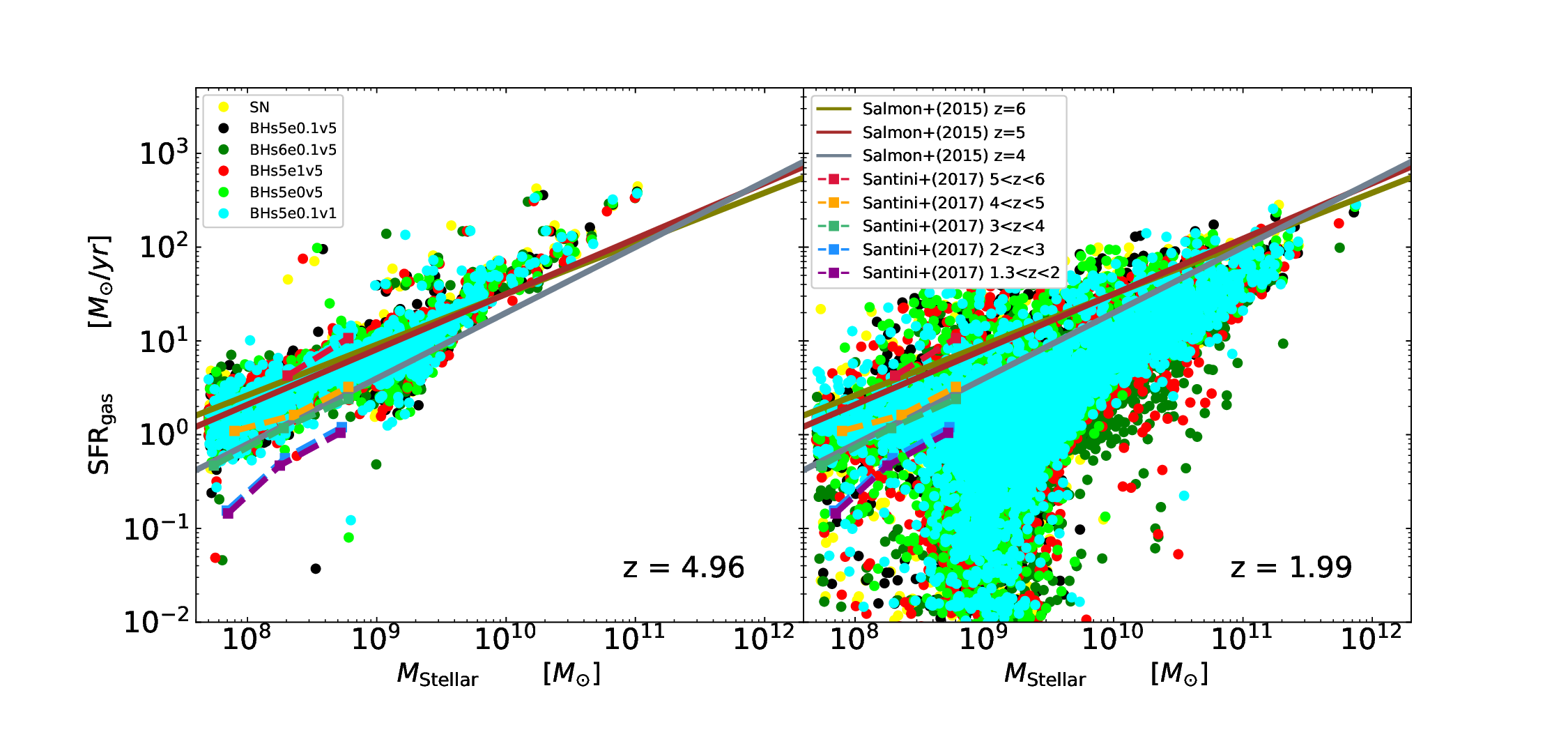} 
\vspace{-1.0cm} 
\caption{ 
SFR versus stellar mass of the galaxies in our simulations, 
at the epoch $z \sim 5$ in the left, and $z \sim 2$ in the right panel. 
The filled circles are our simulation results, 
and the plotting colours distinguish results from different runs. 
The lines indicate observational results of SFR versus stellar mass relation for: 
star-forming galaxies at $z = 4, 5, 6$ \citep{Salmon15} as the solid lines, and 
galaxies in different redshift bins within $z = 1.3 - 6$ \citep{Santini17} 
as the dashed lines with square plotting symbols. 
} 
\label{fig-SFR-vs-StellarMass} 
\end{figure*} 
%%%%%%%%%%%%%%%%%%%%%%%%%%%%%%%%%%%%%%%%%%%%%%%%%%%%%%%%%%%%%%%%%%%%%%%% 

\subsection{Star Formation Rate} 

We compute the Star Formation Rate Density (SFRD) (in $M_{\odot}$ yr$^{-1}$ Mpc$^{-3}$) 
by summing over all the SF occurring in each simulation box at a time, 
and dividing it by the time-step interval and the box volume. 
The global SFRD as a function of redshift is presented in Fig.~\ref{fig-SFRD}, 
the eight simulations labelled by the different colours. 
Observational data limits are shown, for a comparison, as the grey shaded region, 
taken mainly from \citet{Cucciati12}, and the compilations therein originally from 
\citet{PerezGonzalez05}, \citet{Schiminovich05}, \citet{Bouwens09}, 
\citet{Reddy09}, \citet{Rodighiero10}, \citet{vanderBurg10}, \citet{Bouwens12}. 
The SFRD from these observations continues to grow from 
early cosmic epochs at $z < 10$, and has a peak around $z \sim 2 - 3$. 

% Gas reservoirs in galaxies are being replenished by cosmic accretion and galaxy mergers. 

In our simulations we find that the SFRD rises with time from early epochs $z \sim 10$, 
reach a maximum SFRD in the form of a wide peak at $z \sim 2-3$, 
and the SFRD decreases at $z < 3$; an overall trend consistent with the observations. 
Star formation occurs inside galaxies, where cosmic large-scale-structure gas inflows and cools. 
The presence of a central SMBH helps to quench SF, 
because a fraction of gas is ejected out and/or heated by BH feedback, 
and a small fraction might be accreted onto the BH. 

% The SFRD in the simulation {\it BHs4e0.1v5} (magenta curve) is almost similar to that 
% in the run {\it SN}, because the BHs are too small there to generate enough feedback. 
% A similar outcome happens in the other runs at $z \geq 4$, when the BHs are too small. 

We consider the {\it SN} run (yellow curve) without BHs as the baseline, 
and compare other simulations with it to estimate the impact of BH feedback. 
Note that in the run {\it BHs5e0v5} (light green curve) no BH feedback is implemented, 
or $\epsilon_f = 0$; and here the SF remains the same as the {\it SN} run. 
However growth of the BHs by gas accretion and mergers is present. 
Thus we conclude that gas removal by accretion onto BHs play only a very minor role in suppressing SF, 
and the major role is played by BH feedback induced removal/heating of gas. 

We find that the physical processes of BH accretion and feedback 
causes a quenching of the SFRD compared to the {\it SN} case 
(yellow curve in Fig.~\ref{fig-SFRD}), at cosmic epochs $z \leq 4$. 
The reduction of SFRD factors at $z = 0$ 
for the different simulations range between $1.3 - 15$ times. 

% Thus, we find that BHs need to grow to a mass $M_{\rm BH} \sim 10^8 M_{\odot}$ or more massive, 
% to generate enough feedback energy in order to suppress star-formation in our simulated galaxies. 

To explore the galaxy main-sequence, we plot the 
SFR (in $M_{\odot}/yr$) versus stellar mass ($M_{\star}$) of all the galaxies 
within the cosmological boxes in Fig.~\ref{fig-SFR-vs-StellarMass}; 
at the epoch $z \sim 5$ in the left panel, and $z \sim 2$ at the right. 
The filled circles are our simulation results, with the plotting colours 
distinguishing results from different runs. 
The solid lines indicate observational data from \citet{Salmon15} 
of star-forming galaxies at $z = 4, 5, 6$: the best-fit relation of SFR versus 
$M_{\star}$ as written in their Equation (6) with parameters given in their Table 4. 
The dashed lines show observations from \citet{Santini17} in different redshift bins 
within $z = 1.3 - 6$: the data points are taken from their Table 1. 

We find that our simulated galaxies are well consistent with the $[$SFR $- M_{\star}]$ relations 
showing the observational main sequence of star-forming galaxies. 
There is however a large scatter in our SFR versus $M_{\star}$ correlation.

\section{Summary and Conclusions} 
\label{sec-conclusion} 

How the central supermassive black holes grew to billions of solar masses 
and produced the varied populations of AGN that we observe, involves unresolved questions. 
Especially at early epochs, luminous $z \sim 6$ quasars are observed to host 
$10^9 M_{\odot}$ SMBHs when the Universe was $< 1$ Gyr old. 
It is challenging to understand how they formed over such short time-scales, 
with varying possible theories: 
growth from stellar-mass seed BHs by rapid enhanced super-Eddington accretion, 
formation of direct-collapse heavy $10^5 M_{\odot}$ BH seeds soon after the Big Bang, 
or via mergers of intermediate-mass black holes. 

Adding to the population of early SMBHs, recent JWST observations are detecting 
many SMBHs at $z \sim 4-12$ which are overmassive in their host galaxy, 
as well as some SMBHs which are normal-mass with respect to the local $M_{\rm BH} - M_{\star}$ correlation. 
The exact formation channels of these diverse populations of 
normal-mass and overmassive SMBHs are unknown. 

We have investigated the growth of SMBHs versus host galaxies, 
as well as their feedback, by performing cosmological hydrodynamical simulations. 
Using a modified version of the SPH code GADGET-3, 
we simulated $(50 ~ {\rm Mpc})^3$ comoving volumes, 
with a mass resolution of $4.61 \times 10^{7} M_{\odot}$ for gas particles, 
from $z = 100$ up to $z = 0$. 
The simulations include the sub-resolution physics of radiative cooling, star-formation, 
stellar evolution, chemical enrichment, SN feedback, AGN accretion and feedback. 
We probe the BH-galaxy co-evolution 
in terms of the black hole mass - stellar mass correlation. 

We executed a series of simulations: two of them are control cases; 
one with SF-only and one additionally SN feedback; the other runs include BHs as well. 
We explore different parameter variations of the BH sub-resolution models: 
$M_{\rm BHseed} = (10^{4}, 10^{5}, 10^{6}) M_{\odot}$, 
$\epsilon_f = 0, 0.1, 1$, 
the outflow velocity for BH kinetic feedback: $v_w = 1000, 5000, 10000$ km/s. 

Based on our simulations, in our study we find the following: 

\begin{itemize} 

\item The initial growth from $10^5 M_{\odot}$ seeds to $\sim 10^7 M_{\odot}$ BHs 
occurs predominantly via BH mergers. 

\item Gas accretion onto the BHs is initially low, with highly sub-Eddington 
accretion rates ($\dot{M}_{\rm BH} / \dot{M}_{\rm Edd} < 0.001$). 
$\dot{M}_{\rm BH}$ increases with time, and reaches the Eddington rate after $7-9$ Gyrs. 
The BHs then undergo very fast growth via Eddington-limited 
(or Eddington ratio $= 1$) gas accretion. 
Within a period of $600 - 700$ Myr, 
the BHs grow from $10^7 M_{\odot}$ to $(10^9 - 10^{10}) M_{\odot}$. 

\item Supermassive BHs (those reaching $10^9 - 10^{10} M_{\odot}$ at $z=0$) 
have had their predominant growth over a period of $600 - 700$ Myr via strong gas accretion. 
During this period, the BH mass increases by a factor $10^2 - 10^3$. 
We argue that such BH growth behavior results from the influence of 
gas supply to the host galaxy center, and subsequent efficient gas accretion onto the BH; 
which leads to central gas depletion within $700$ Myr. 

\item After the Eddington-limited gas accretion growth to $\sim 10^9 - 10^{10} M_{\odot}$, 
the SMBHs stop growing and maintains the same mass up to $z = 0$. 
This final growth halt happens because of gas depletion from galaxy centers. 

\item The most-massive population of BHs 
have grown to $M_{\rm BH} \sim 10^{10} M_{\odot}$ at $z = 0$. 
BHs do not grow to more than $10^{10} M_{\odot}$, 
because of gas removal by AGN feedback driven self-regulation. 
Some BHs may reach this maximum mass 2 or 3 Gyr ago. 

\item Applying our results to the JWST detected high-$z$ Little Red Dots, we argue that: 
the normal-mass SMBHs had predominantly undergone BH merger driven evolution, 
whereas the overmassive BHs underwent periods of Eddington-limited 
or super-Eddington bursts of gas accretion. 

\item Our simulations probe galaxies with a 
stellar mass between $M_{\star} = (10^{8} - 10^{12}) M_{\odot}$. 
The star formation rate density ($M_{\odot}$ yr$^{-1}$ Mpc$^{-3}$) versus redshift evolution, 
as well as the main-sequence of SFR-stellar mass relation of these galaxies, 
are consistent with observations. 
Star-formation is quenched at $z = 3 - 0$, and the SFRD is reduced by factors $1.3 - 15$. 

\item Gas removal by accretion onto BHs play only a very minor role in suppressing SF, 
and the major role is played by BH feedback induced removal/heating of gas. 

\end{itemize} 

{\it 
We deduce that supermassive $10^9 - 10^{10} M_{\odot}$ BHs have had their predominant growth 
over a period of $600 - 700$ Myr via efficient gas accretion, 
from the gas supply to the host galaxy center, 
when the BH mass increases $10^2 - 10^3$ times. 
BH mergers play a minor role compared to gas accretion for SMBH growth. 
SMBHs do not grow to more than $10^{10} M_{\odot}$, 
because of gas depletion from galaxy centers driven by AGN feedback. 
}

%%%%%%%%%%%%%%%%   REFERENCES


\bibliographystyle{aa} % style aa.bst

\bibliography{reference_Papers} % your references Yourfile.bib

\section{Acknowledgements} 
P.B. is most grateful to Volker Springel and Klaus Dolag for allowing to use the GADGET-3 code. 
P.B. acknowledges useful discussions with Giuseppe Murante, Matteo Viel, Luca Tornatore, 
Stefano Borgani, Gian Luigi Granato, Pierluigi Monaco, Simona Gallerani, and Andrea Ferrara. 
The work has been supported by the Brazilian funding Agency FAPESP (grants 2016/01355-5 and 2016/22183-8); 
and the Italian Ministry of University and Research (MUR) 
Missione 4 "Istruzione e Ricerca" - Componente C2, 
Investimento 1.1 Fondo per il Programma Nazionale di Ricerca e Progetti di Rilevante Interesse Nazionale (PRIN), 
the PRIN 2022 PNRR grant under the National Recovery and Resilience Plan (PNRR): 
project P2022ZLW4T 
"Next-generation computing and data technologies to probe the cosmic metal content".

% %%%%%%%%%%%%%%%%%%%%%%%%%%%%%%%%%%%%%%%%%%%%%%%%%%%%%%%%%%%%%%%%%%%%%%%% 
% % FIGURE A.1 
% \begin{figure*} 
% \centering 
% \includegraphics[width = 1.07 \linewidth]{../../Plots/GFE-vs-HaloMass/GFE_vs_HaloDMmass.eps} 
% \vspace{-2cm} 
% \caption{ 
% GFE versus DM halo mass of the galaxies within the simulation volume, 
% at four cosmic epochs $z = 7.94$ in the left, and $z = 5.49$ in the right panel. 
% The filled circles are our simulation results, with the plotting colours 
% distinguishing results from different runs. 
% } 
% \label{fig-GFE-vs-HaloDMmass} 
% \end{figure*} 
% %%%%%%%%%%%%%%%%%%%%%%%%%%%%%%%%%%%%%%%%%%%%%%%%%%%%%%%%%%%%%%%%%%%%%%%% 

%%%%%%%%%%%%%%%%%%%%%%%%%%%%%%%%%%%%%%%%%%%%%%%%%%%%%%%%%%%%%%% 

\end{document}